\documentclass[12pt]{article}
  \usepackage{amsfonts}
  \usepackage{amsmath}
\usepackage{amssymb}
\usepackage{amscd}
\usepackage{graphicx}
\begin{document}
~~
\bigskip
\bigskip
\begin{center}
{\Large {\bf{{{The Henon-Heiles system defined on canonically deformed space-time}}}}}
\end{center}
\bigskip
\bigskip
\bigskip
\begin{center}
{{\large ${\rm {Marcin\;Daszkiewicz}}$ }}
\end{center}
\bigskip
\begin{center}
{ {{{Institute of Theoretical Physics\\ University of Wroc{\l}aw pl.
Maxa Borna 9, 50-206 Wroc{\l}aw, Poland\\ e-mail:
marcin@ift.uni.wroc.pl}}}}
\end{center}
\bigskip
\bigskip
\bigskip
\bigskip
\bigskip
\bigskip
\bigskip
\bigskip
\begin{abstract}
In this article we provide canonically deformed classical Henon-Heiles system. Further we demonstrate that for proper value of
deformation parameter $\theta$ there appears chaos in the model.
\end{abstract}
\bigskip
\bigskip
\bigskip
\bigskip
\bigskip
\bigskip
\bigskip
\bigskip
\bigskip
 \eject
\section{{{Introduction}}}

There exist a lot of papers dealing with physical models of which dynamics remains chaotic; the most popular of them are:  Lorentz system
 \cite{lorenc}, Henon-Heiles system  \cite{henon}, Rayleigh-Bernard system \cite{rbsystem},
Duffing equation \cite{duffing}, Double pendulum \cite{dpendulum}, \cite{dpendulum1},  Forced damped
pendulum \cite{dpendulum}, \cite{dpendulum1} and
Quantum forced damped oscillator model \cite{quantumdamped}. The especially interesting seems to be  Henon-Heiles system defined by the following Hamiltonian function
\begin{eqnarray}
H(p,x) \;=\; \frac{1}{2}\sum_{i=1}^2\; \left(p_i^2 + x_i^2\right) + x_1^2x_2 - \frac{1}{3}x_2^3\;,\label{ham}
\end{eqnarray}
which in cartesian coordinates $x_1$ and $x_2$ describes the set of two nonlinearly coupled harmonic oscillators.
In polar coordinates $r$ and $\theta$ it corresponds to the particle moving in noncentral potential of the form
\begin{eqnarray}
V(r,\varphi) \;=\; \frac{r^2}{2} + \frac{r^3}{3}\sin\left(3\varphi\right)\;,\label{pot}
\end{eqnarray}
with $x_1 = r\cos\varphi$ and $x_2 = r\sin\varphi$. The above model has been inspired by the observational data indicating, that star moving
in a weakly  perturbated central potential should has apart of constant in time total energy $E_{\rm{tot}}$, the second
conserved physical quantity $I$. It has been demonstrated with use of so-called
Poincare section method, that such a situation appears in the case of Henon-Heiles system only for small values of control parameter
$E_{\rm{tot}}$. For high energies the trajectories in phase space become chaotic and the quantity $I$ does not exist (see e.g. \cite{tabor},
\cite{genhamhh}).

In this article we investigate the impact of the well-known (simplest) canonically deformed Galilei space-time \cite{oeckl}-\cite{dasz1}\footnote{ The canonically noncommutative space-times have been  defined as the quantum
representation spaces, so-called Hopf modules (see e.g. \cite{oeckl}, \cite{chi}), for the canonically deformed quantum Galilei Hopf algebras $\,{\mathcal U}_\theta({ \cal G})$.},\footnote{It should be noted that in accordance with the
Hopf-algebraic classification  of all deformations of relativistic
and nonrelativistic symmetries (see references \cite{class1}, \cite{class2}),
apart of canonical \cite{oeckl}-\cite{dasz1} space-time noncommutativity, there also exist  Lie-algebraic \cite{dasz1}-\cite{lie1} and quadratic \cite{dasz1}, \cite{lie1}-\cite{paolo} type of quantum spaces.}
\begin{equation}
[\;t,\hat{x}_{i}\;] = 0\;\;\;,\;\;\; [\;\hat{x}_{i},\hat{x}_{j}\;] = 
i\theta_{ij}
\;,\label{nhspace}
\end{equation}
on the mentioned above Henon-Heiles system. Particularly, we provide the corresponding canonical equations of motion as well as we find the Poincare sections of the phase space trajectories of
the model. In such a way we demonstrate that for proper value of deformation parameter $\theta$ and for proper values of
control parameter $E_{\rm{tot}}$ there appears chaos. 

The paper is organized as follows. In second Section we briefly remaind the basic properties of classical Henon-Heiles system. In Section 3 we recall
canonical noncommutative quantum Galilei space-time proposed in article \cite{dasz1}. Section 4 is devoted to the new canonically deformed  Henon-Heiles model while the conclusions are discussed in Final remarks.

\section{{{Classical Henon-Heiles model}}}

As it was already mentioned the Henon-Heiles system is defined by the following Hamiltonian function
\begin{eqnarray}
H(p,x) \;=\; \frac{1}{2}\sum_{i=1}^2\; \left(p_i^2 + x_i^2\right) + x_1^2x_2 - \frac{1}{3}x_2^3\;,\label{ham1}
\end{eqnarray}
with canonical variables $(p_i,x_i)$ satisfying
\begin{equation}
\{\;x_i,x_j\;\} = 0 =\{\;p_i,p_j\;\}\;\;\;,\;\;\; \{\;x_i,p_j\;\}
=\delta_{ij}\;, \label{classpoisson}
\end{equation}
i.e., it describes the system of two nonlineary coupled one-dimensional harmonic oscillator models. One can check that the corresponding
canonical equations of motion take the form
\begin{eqnarray}
&& \dot{p}_1\;=\; -\frac{\partial H}{\partial x_1}\;=\; -x_1 -2x_1x_2\;\;\;,\;\;\;
\dot{x}_i\;=\; \frac{\partial H}{\partial p_i}\;=\; p_i\;,\label{eqs2}\\[5pt]
&&\dot{p}_2\;=\; -\frac{\partial H}{\partial x_2}\;=\; -x_2 -x_1^2 + x_2^2 \;,
\label{eqs3}
\end{eqnarray}
while the proper Newton equations look as follows
\begin{equation}
\left\{\begin{array}{rcl}
\ddot{x}_1&=& -x_1 - 2x_1x_2\\[5pt]
\ddot{x}_2&=&  -x_2 -x_1^2 + x_2^2 \;.
\end{array}\right.\label{neqs2}
\end{equation}
Besides, it is easy to see that the conserved in time total energy of the model is given by
\begin{eqnarray}
E_{\rm tot} \;=\; \frac{1}{2}\sum_{i=1}^2\; \left(\dot{x}_i^2 + x_i^2\right) + x_1^2x_2 - \frac{1}{3}x_2^3\;.\label{energy}
\end{eqnarray}

In order to analyze the discussed system we find numerically the
Poincare maps in two dimensional phase space $(x_2,p_2)$ for section $x_1 = 0$ and for five fixed values of total energy: $E_{\rm tot} = 0.03125$, $E_{\rm tot} = 0.06125$, $E_{\rm tot} = 0.10125$, $E_{\rm tot} = 0.125$, $E_{\rm tot} = 0.15125$ and $E_{\rm tot} = 0.16245$ respectively; the obtained results are summarized on {\bf Figures 1 - 6}. We see that for $E_{\rm tot} = 0.03125$ the trajectories remain completely regular. However, for increasing values of control parameter $E_{\rm tot}$ they gradually become disordered until to the almost completely chaotic behavior of the system at $E_{\rm tot} = 0.16245$\footnote{The calculations are performed for single trajectory with initial condition $x_1(0) = \left({2}{E_{\rm tot}}\right)^{\frac{1}{2}}$ and $x_2(0) = p_1(0) = p_2(0) = 0$.}.

\section{Canonically deformed Galilei space-time}

In this section we very shortly recall the basic facts associated with the (twisted) canonically deformed  Galilei Hopf algebra
$\;{\cal U}_{\theta}({\cal G})$ and with the
corresponding quantum space-time \cite{dasz1}.  Firstly, it should be noted that in accordance with Drinfeld  twist procedure \cite{drin} the algebraic sector of
Hopf structure $\;{\cal U}_{\theta}({\cal G})$ remains
undeformed
\begin{eqnarray}
&&\left[\, K_{ij},K_{kl}\,\right] =i\left( \delta
_{il}\,K_{jk}-\delta
_{jl}\,K_{ik}+\delta _{jk}K_{il}-\delta _{ik}K_{jl}\right) \;,  \label{ff} \\
&~~&  \cr &&\left[\, K_{ij},V_{k}\,\right] =i\left( \delta
_{jk}\,V_i-\delta _{ik}\,V_j\right)\;\; \;, \;\;\;\left[
\,K_{ij},\Pi_{\rho }\right] =i\left( \eta _{j \rho }\,\Pi_{i }-\eta
_{i\rho }\,\Pi_{j }\right) \;, \label{nnn}
\\
&~~&  \cr &&\left[ \,V_i,V_j\,\right] = \left[ \,V_i,\Pi_{j
}\,\right] =0\;\;\;,\;\;\;\left[ \,V_i,\Pi_{0 }\,\right]
=-i\Pi_i\;\;\;,\;\;\;\left[ \,\Pi_{\rho },\Pi_{\sigma }\,\right] =
0\;,\label{ff1}
\end{eqnarray}
where $K_{ij}$, $\Pi_0$, $\Pi_i$ and $V_i$ can be identified with  rotation, time translation, momentum and boost operators respectively. Besides,
the coproducts and antipodes of such  algebra take the form
\begin{eqnarray}
&&\Delta_\theta(\Pi_\rho)=\Delta_0(\Pi_\rho)\;\;\;,\;\;\;
\Delta _{\theta }(V_i) =\Delta _{0}(V_i)\;, \label{dlww3v}\\
&~~~&  \cr \Delta _{\theta }(K_{ij})
&=&\Delta _{0}(K_{ij})-%
\theta ^{k l }[(\delta_{k i}\Pi_{j }-\delta_{k j
}\,\Pi_{i})\otimes \Pi_{l }\nonumber\\
&&\qquad\qquad\qquad\qquad\qquad+\Pi_{k}\otimes (\delta_{l
i}\Pi_{j}-\delta_{l j}\Pi_{i})]\;,\label{zadruzny}\\
&~~~&  \cr
&& S(\Pi_{\rho})=-\Pi_{\rho}\;\;\;,\;\;\;S(K_{ij})=-K_{ij}\;\;\;,\;\;\;S(V_i)=-V_i\;,  \label{zadruga}
\end{eqnarray}
 while the corresponding quantum space-time can be defined as the representation space, so-called Hopf modules (see e.g. \cite{oeckl}, \cite{chi}), for the canonically deformed Hopf structure
$\;{\cal U}_{\theta}({\cal G})$; it looks as follows
\begin{equation}
[\;t,\hat{x}_{i}\;] = 0\;\;\;,\;\;\; [\;\hat{x}_{i},\hat{x}_{j}\;] = 
i\theta_{ij}
\;,\label{nhspace1}
\end{equation}
and for  deformation parameter $\theta$  approaching  zero it 
becomes commutative.

\section{{{Classical Henon-Heiles system on canonically deformed space-time}}}

Let us now turn to the Henon-Heiles model defined on quantum space-time (\ref{nhspace1}).
In first step of our construction we extend the canonically deformed space to the whole algebra of momentum
and position operators as follows
(see e.g. \cite{chaihydro1}-\cite{romero1})\footnote{The correspondence relations are $\{\;\cdot,\cdot\;\} = \frac{1}{i}\left[\;\cdot,\cdot\;\right]$.}
\begin{eqnarray}
&&\{\;\hat{ x}_{1},\hat{ x}_{2}\;\} = 2\theta\;\;\;,\;\;\;
\{\;\hat{ p}_{i},\hat{ p}_{j}\;\} = 0\;\;\;,\;\;\; \{\;\hat{ x}_{i},\hat{ p}_{j}\;\} = \delta_{ij}\;.\label{rel1}
\end{eqnarray}
 One can check that relations
(\ref{rel1}) satisfy the Jacobi identity and for deformation parameter
$\theta$ approaching zero become classical. \\
Next, by analogy to the commutative case we define the corresponding Hamiltonian function by\footnote{Such a construction of deformed Hamiltonian
function (by replacing the commutative variables $(x_i,p_i)$ by noncommutative ones $({\hat x}_i, {\hat p}_i)$) is well-known in the
literature - see e.g. \cite{chaihydro1}, \cite{chaihydro2} and \cite{exem}.}
\begin{eqnarray}
H(\hat{p},\hat{x}) \;=\; \frac{1}{2}\sum_{i=1}^2\; \left(\hat{p}_i^2 + \hat{x}_i^2\right) + \hat{x}_1^2\hat{x}_2 - \frac{1}{3}\hat{x}_2^3\;,\label{nonham}
\end{eqnarray}
with the
noncommutative operators $({\hat x}_i, {\hat p}_i)$  represented by the classical
ones $({ x}_i, { p}_i)$ as  \cite{romero1}-\cite{kijanka}
\begin{eqnarray}
{\hat x}_{1} &=& { x}_{1} - { \theta}p_2\;,\label{rep1}\\[5pt]
{\hat x}_{2} &=& { x}_{2} +{\theta}p_1
\;,\label{rep2}\\[5pt]
{\hat p}_{1} &=& { p}_{1}\;\;,\;\;{\hat p}_{2} \;=\; { p}_{2}\;.\label{rep2a}\\[5pt]
\end{eqnarray}
Consequently, we have 
\begin{eqnarray}
H({p},{x}) &=&
\frac{1}{2M(\theta)}\left({{{p}}_1^2}+{{{p}}_2^2} \right)  +
\frac{1}{2}M(\theta)\Omega^2({\theta})\left({{{x}}_1^2}+{{{x}}_2^2} \right)
- S(\theta)L\;+ \label{2dh1}\\
&+& \left(x_1 - {{\theta}p_2}\right)^2\cdot \left(x_2 + {{ \theta}p_1}\right)
- \frac{1}{3}\left(x_2 + {{\theta}p_1}\right)^3\;,\nonumber
\end{eqnarray}
where
\begin{eqnarray}
&&L = x_1p_2 - x_2p_1\;, \\[5pt]
&&1/M({\theta}) = 1 +{\theta}^2 \;,\\[5pt]
&&\Omega({\theta}) = \sqrt{\left(1
+{\theta}^2 \right)}\;,
\end{eqnarray}
and
\begin{eqnarray}
S({\theta})={\theta}\;.
\end{eqnarray}
Further, using the 
formula (\ref{2dh1}) one gets the following canonical Hamiltonian equations
of motion
\begin{eqnarray}
\dot{x}_1 &=& \frac{1}{M({\theta})}p_1 + S({\theta})x_2 +
\left[\left(x_1- {\theta}p_2\right)^2 - \left(x_2+ {\theta}p_1\right)^2\right]{\theta}\;,\\[5pt]
\dot{x}_2 &=& \frac{1}{M({\theta})}p_2 - S({\theta})x_1 - 2\left(x_2+ {\theta}p_1\right)\left(x_1- {\theta}p_2\right){\theta}\;,\\[5pt]
\dot{p}_1 &=& -{M({\theta})}\Omega^2({\theta})x_1 + S({\theta})p_2 -
2\left(x_2+ {\theta}p_1\right)\left(x_1- {\theta}p_2\right)\;,\\[5pt]
\dot{p}_2 &=& -{M({\theta})}\Omega^2({\theta})x_2 + S({\theta})p_1 - \left(x_1- {\theta}p_2\right)^2 +
\left(x_2+ {\theta}p_2\right)^2\;,
\end{eqnarray}
which for deformation parameter running to zero become classical.


Similarly to the undeformed case we
find numerically the
Poincare maps in two dimensional phase space $(x_2,p_2)$ for section $x_1=0$.  However, this time apart of parameter
$E_{\rm tot}$ we take under consideration the parameter of deformation $\theta$. Consequently, we derive the Poincare sections of phase space parameterized
by pair $\left(E_{\rm tot}, \theta\right)$ for $\theta = 0.5,1,2$ and six values of total energy
$E_{\rm tot}$. In such a way we detect chaos in the model only for $\theta = 0.5$ and for $E_{\rm tot} = 0.160178$, $E_{\rm tot} = 0.1607445$ and
$E_{\rm tot} = 0.16245$
respectively (see for chaotic scenario {\bf Figures 7 - 12}).
In the case $\theta=1$ as well as $\theta =2$ the system remains ordered\footnote{As in the undeformed case the calculations are performed for single trajectory with initial condition $x_1(0) = \left({2}{E_{\rm tot}}\right)^{\frac{1}{2}}$ and $x_2(0) = p_1(0) = p_2(0) = 0$.}.

\section{{{Final remarks}}}

In this article we provide the canonically deformed Henon-Heiles system, i.e., we define the proper Hamiltonian function as well as we derive the corresponding equations of motion. We also demonstrate (with use of the Poincare section method) that for deformation
parameter $\theta=0.5$ and for particular values of control parameter $E_{\rm{tot}}$ the analyzed model becomes chaotic.

As a next step of presented here investigations one can consider the canonical deformation of so-called generalized Henon-Heiles systems given by the following Hamiltonian function
\begin{eqnarray}
H(p,x) \;=\;  \frac{1}{2} \left(p_1^2 + p_2^2\right) + \delta x_1^2 + (\delta+\Omega)x_2^2 + \alpha x_1^2x_2 +\alpha\beta x_2^3\;,\label{genhamh}
\end{eqnarray}
with arbitrary coefficients $\alpha$, $\beta$, $\delta$ and $\Omega$ respectively. It should be noted that the properties of commutative models described by function (\ref{genhamh})
are quite interesting. For example, it is well-known (see e.g. \cite{p1}-\cite{p10} and references therein) that such systems remain integrable only in the Sawada-Kotera case: with $\beta =1/3$ and
$\Omega = 0$, in the KdV case: with $\beta =2$ and arbitrary  $\Omega$ as well as in the Kaup-Kupershmidt case: with $\beta = 16/3$ and $\Omega =15\delta$. Besides there has been provided in articles \cite{bal1} and \cite{bal2}\footnote{See also references therein.} the different types of integrable perturbations of mentioned above (integrable) models such as, for example, $q^{-2}$ perturbations, the Ramani series of polynomial deformations and the rational perturbations.  Consequently the impact of the canonical deformation (\ref{nhspace}) on the above dynamical structures (in fact) seems to be very interesting. For this reason
the works in this direction already started and are in progress.

\section*{Acknowledgments}
The author would like to thank J. Lukierski and K. Graczyk
for valuable discussions.\\
This paper has been financially supported by Polish Ministry of
Science and Higher Education grant NN202318534.

\eject

$~~~~~~~~~~~~~~~~~~$
\\
\\
\\
\\
\\
\begin{figure}[htp]
\includegraphics[width=\textwidth]{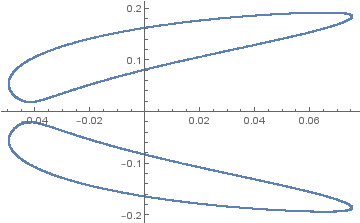}
\caption{Poincare map in two dimensional phase space $(x_2,p_2)$ for section $x_1 = 0$ and for total energy $E_{\rm tot} = 0.03125$. Trajectory is completely regular - there is no chaos in the system.}\label{rysunek1}
\end{figure}
\eject

$~~~~~~~~~~~~~~~~~~$
\\
\\
\\
\\
\\
\begin{figure}[htp]
\includegraphics[width=\textwidth]{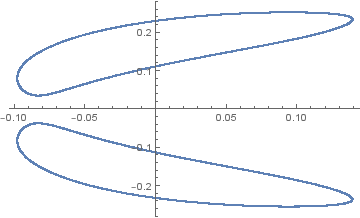}
\caption{Poincare map in two dimensional phase space $(x_2,p_2)$ for section $x_1 = 0$ and for fixed value of total energy $E_{\rm tot} = 0.06125$. Trajectory is still regular - the system is chaos free.}\label{rysunek2}
\end{figure}
\eject

$~~~~~~~~~~~~~~~~~~$
\\
\\
\\
\\
\\
\begin{figure}[htp]
\includegraphics[width=\textwidth]{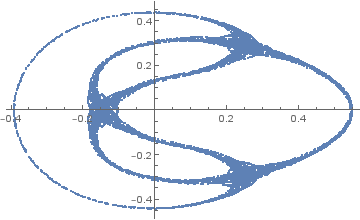}
\caption{Poincare map in two dimensional phase space $(x_2,p_2)$ for section $x_1= 0$ and for total energy $E_{\rm tot} = 0.10125$. Trajectory remains still regular.}\label{rysunek3}
\end{figure}
\eject

$~~~~~~~~~~~~~~~~~~$
\\
\\
\\
\\
\\

\begin{figure}[htp]
\includegraphics[width=\textwidth]{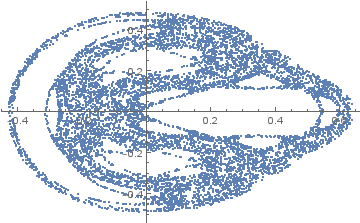}
\caption{Poincare map in two dimensional phase space $(x_2,p_2)$ for section $x_1= 0$ and for total energy $E_{\rm tot} = 0.125$. The system becomes mixed:
chaotic and ordered simultaneously.}\label{rysunek4}
\end{figure}
\eject

$~~~~~~~~~~~~~~~~~~$
\\
\\
\\
\\
\\
\begin{figure}[htp]
\includegraphics[width=\textwidth]{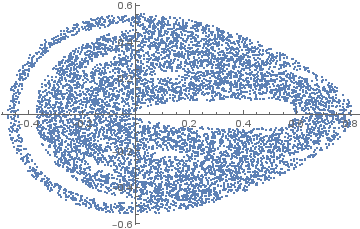}
\caption{Poincare map in two dimensional phase space $(x_2,p_2)$ for section $x_1= 0$ and for total energy $E_{\rm tot} = 0.15125$. The system becomes chaotic.}\label{rysunek5}
\end{figure}
\eject

$~~~~~~~~~~~~~~~~~~$
\\
\\
\\
\\
\\
\begin{figure}[htp]
\includegraphics[width=\textwidth]{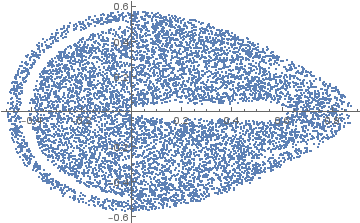}
\caption{Poincare map in two dimensional phase space $(x_2,p_2)$ for section $x_1= 0$ and for total energy $E_{\rm tot} = 0.16245$. The chaos
increases.}\label{rysunek6}
\end{figure}

\eject

$~~~~~~~~~~~~~~~~~~$
\\
\\
\\
\\
\\
\begin{figure}[htp]
\includegraphics[width=\textwidth]{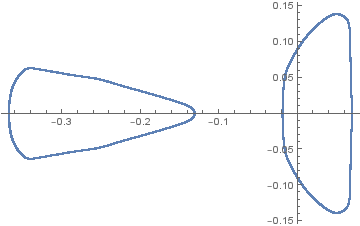}
\caption{Poincare map in two dimensional phase space $(x_2,p_2)$ for section $x_1 = 0$ and for total energy $E_{\rm tot} = 0.15125$. Trajectory is completely regular - there is no chaos in the system.}\label{rysunek7}
\end{figure}
\eject

$~~~~~~~~~~~~~~~~~~$
\\
\\
\\
\\
\\
\begin{figure}[htp]
\includegraphics[width=\textwidth]{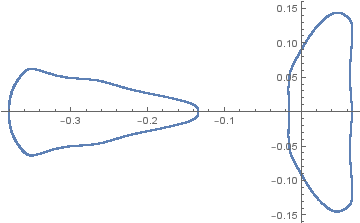}
\caption{Poincare map in two dimensional phase space $(x_2,p_2)$ for section $x_1 = 0$ and for fixed value of total energy $E_{\rm tot} = 0.1568$. Trajectory is still regular - the system is chaos free.}\label{rysunek8}
\end{figure}
\eject

$~~~~~~~~~~~~~~~~~~$
\\
\\
\\
\\
\\
\begin{figure}[htp]
\includegraphics[width=\textwidth]{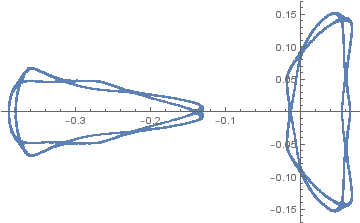}
\caption{Poincare map in two dimensional phase space $(x_2,p_2)$ for section $x_1= 0$ and for total energy $E_{\rm tot} = 0.1596125$. Trajectory remains still regular.}\label{rysunek9}
\end{figure}
\eject

$~~~~~~~~~~~~~~~~~~$
\\
\\
\\
\\
\\
\begin{figure}[htp]
\includegraphics[width=\textwidth]{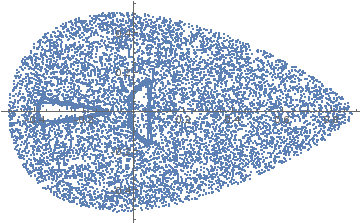}
\caption{Poincare map in two dimensional phase space $(x_2,p_2)$ for section $x_1= 0$ and for total energy $E_{\rm tot} = 0.160178$. The system becomes suddenly
chaotic.}\label{rysunek10}
\end{figure}
\eject

$~~~~~~~~~~~~~~~~~~$
\\
\\
\\
\\
\\
\begin{figure}[htp]
\includegraphics[width=\textwidth]{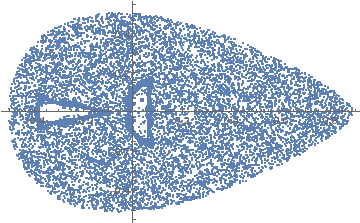}
\caption{Poincare map in two dimensional phase space $(x_2,p_2)$ for section $x_1= 0$ and for total energy $E_{\rm tot} = 0.1607445$.
 The chaos increases.}\label{rysunek11}
\end{figure}
\eject

$~~~~~~~~~~~~~~~~~~$
\\
\\
\\
\\
\\
\begin{figure}[htp]
\includegraphics[width=\textwidth]{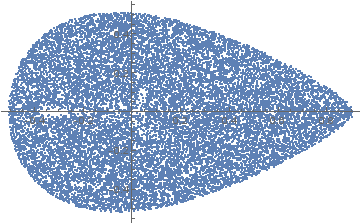}
\caption{Poincare map in two dimensional phase space $(x_2,p_2)$ for section $x_1= 0$ and for total energy $E_{\rm tot} = 0.16245$. The system becomes totally
chaotic.}\label{rysunek12}
\end{figure}

\end{document}